\documentclass{amsart}
\usepackage[utf8]{inputenc}
\usepackage[T1]{fontenc}
\usepackage{cite}
\usepackage{hyperref}
\usepackage{graphicx}
\usepackage{amsaddr}
\usepackage{setspace}
\title[Optimizing information flow in GRNs]{
Optimizing information flow in Gene Regulatory Networks: a geometric perspective}

\author[Bravetti et al.]{Alessandro Bravetti$^1$, Miguel Ángel García Ariza$^{2,\star}$, Pablo Padilla-Longoria$^2$, José Roberto Romero-Arias$^2$}
\address{$^1$School of Science and Technology, University of Camerino, Camerino, Italy}
\address{$^2$Instituto de Investigaciones en Matemáticas Aplicadas y en Sistemas,\\ Universidad Nacional Autónoma de México, Mexico City, Mexico}
\thanks{$\star$\texttt{miguel.garcia@iimas.unam.mx}}

\keywords{Signaling networks, information theory, noise, optimization}

\begin{document}

\begin{abstract}
The dynamics of gene regulatory networks (GRNs) is governed by the interaction between deterministic biochemical reactions and molecular noise. To understand how gene regulatory networks process information during cell state transitions, we study stochastic dynamics derived from a Boolean network model via its representation on the parameter space of Gaussian distributions, equipped with the Fisher information metric.
This reformulation reveals that the trajectories of optimal information transfer are gradient flows of the Kullback-Leibler divergence. We demonstrate that the most efficient dynamics require isotropic decay rates across all nodes and that the noise intensity quantitatively determines the potential differentiation between the initial and final states. Furthermore, we show that paths minimizing biological cost correspond to metric geodesics that require noise suppression, leading to biologically irrelevant deterministic dynamics. Our approach frames noise and decay rates as fundamental control parameters for cellular differentiation, providing a geometric principle for the analysis and design of synthetic networks.
\end{abstract}
\maketitle

\section{Introduction}

In biochemistry, a useful way to think about molecular relationships is in terms of interaction networks. These provide a simplified and powerful representation of cellular systems, capturing relationships between metabolites, proteins, and mRNAs. Among these, gene regulatory networks (GRNs) stand out as key models for understanding gene expression. They are depicted as directed graphs whose dynamics can be described in different ways, including logical formalisms and chemical kinetics.

To understand how gene regulatory networks process information, we need frameworks that integrate biological processes across multiple scales. Research has shown that morphogenetic and topological constraints play a central role in regulatory mechanisms, transcriptional pathways, and overall system robustness in biological systems \cite{villarreal_general_2012,de2020bi,barrio2013cell,denisse,ayala2023boolean,romero2023multiscale,kitano2004biological,feigin2023grn,alon2019introduction}. Recently, we have seen remarkable advances in constructing GRNs and characterizing their topological features \cite{bizzarri2018systems,carlberg2024gene,gomez2020multidimensional,alon2019introduction}. However, a dynamical description of information transfer has often been overlooked, prompting a reevaluation of theoretical approaches to systems biology \cite{kauffman1992origins,gomez2024art,azpeitia2020signaling,tkacik_information_2025}. This gap motivates interdisciplinary work to develop a mathematical language to quantify the flow, storage, and transformation of information within biological networks across multiple scales where fundamental life processes occur \cite{raser_noise_2005, eldar_functional_2010,azpeitia2017combination,briones2024mathematical,roces2018modeling}.  

 Information theory provides powerful tools for these purposes \cite{cover_elements_2006,uda_application_2020}, with applications in various fields, including neuroscience \cite{amari1995information,islas2020information,chen2019enhancer}, biochemical networks \cite{dehmer2011information,momin2025unlocking}, statistical physics \cite{tanaka2000information,kim_information_2021,ohga_information-geometric_2022}, music \cite{paz2022information}, and artificial intelligence \cite{amari1992information,ensslin2022information}. Particularly relevant is \textit{information geometry} \cite{amari_information_2016}, which offers a geometric approach to characterize the dynamics of complex systems far from equilibrium \cite{ito_stochastic_2018,ibanez_heating_2024}. This framework provides metrics derived from information theory that offer new insights for analyzing networks \cite{ito2016introduction,felice_information_2018,yoshimura2021information}. Specifically in the context of GRNs, these methods can reveal how network topologies shape signal pathway fidelity, system robustness, and phenotypic complexity \cite{tkacik_information_2008,tkacik_optimizing_2009,lecca_how_2023}. 

The aim of this paper is to understand how GRNs process information along stochastic dynamics described by the Fokker-Planck (FP) equation. This approach enables us to quantify the Waddington landscape\cite{waddington2014strategy} close to an attractor, and the biological pathways\cite{villarreal_general_2012,wang_quantifying_2011} for development and differentiation, providing a geometric perspective on cellular decision-making processes. Although existing studies have addressed nonstationary dynamics in GRNs \cite{grzegorczyk_modelling_2011,jia_constructing_2010,zhang_gene_2023},
a geometric characterization of information transfer in GRNs remains relatively unexplored, particularly in terms of stochastic dynamics. Here, we address this problem by applying information geometry to GRNs, translating the stochastic dynamics into a deterministic flow on a finite-dimensional space equipped with a metric tensor. This allows us to characterize the trajectories of optimal information transfer as geodesics, while revealing how noise and decay rates control informational efficiency.

More concretely, specific values of the decay rates and noise intensities drive the network dynamics along geodesics. For Gaussian distributions arising from the solution of the FP equation, these curves coincide with gradient flows of the Kullback-Leibler (KL) divergence \cite{fujiwara_gradient_1995,wada_huygens_2023,wada_onsagers_2025}, and can either maximize the rate of information gain or minimize the distance to equilibrium the fastest. Although metric geodesics (minimizing statistical distance) also exist, our analysis shows that they require the suppression of noise and appear to have limited biological relevance.

To formalize these ideas, we first explain how we represent the stochastic dynamics of the network with a deterministic framework. We then introduce the KL divergence and the Fisher metrics, which allow us to define the gradient flow of the divergence and compare it with the deterministic representation of the Langevin dynamics. Next, we study the evolution of networks that occur along geodesics, including those that minimize the statistical distance between states. Finally, we conclude with remarks and perspectives for future work, supplemented by detailed calculations and deeper geometric explanations in the Supplementary Material (SM).

\section{GRN aproach}

There are two well-known approaches to gene networks: the continuous, whose dynamics is dictated by a system of ODEs emerging from chemical kinetics,
and the discrete, where logic functions determine the dynamics. Here, we consider continuous stochastic dynamics derived from a Boolean network of $N$ nodes. The state of the $i$-th one at time $t+1$ is given by $q_i(t+1)=w_i(q(t))$, where the $w_i$ are defined by logical propositions and $q=(q_1,\ldots,q_N)$ are dichotomic variables representing the expression level of each node.

 We may transform this discrete system into a continuous one using fuzzy propositional calculus: the state of the $i$-th node of the network is given by $0\leq q_i\leq1$ and the functions $w_i(q)$ express the interactions of its regulators, with $q=(q_1,\ldots,q_N)$. This state evolves according to the Langevin equation \cite{villarreal_general_2012}
\begin{equation}\label{eq:Langevin}
\frac{\mathrm{d}q_i}{\mathrm{d}t}=\Theta[w_i(q)]-\alpha_iq_i+\xi_i,
\end{equation}
where $\Theta[w_i]$ is its activation,
a sigmoid function,
$\alpha_i$ its relaxation rate,
and $\xi=(\xi_1,\ldots,\xi_N)$ is uncorrelated white noise.

Our aim is to understand how gene regulatory networks process information along these dynamics. Specifically, using tools from information geometry, we study how noise levels and decay rates together optimize information transfer toward steady states. In the process, we also offer a principled framework that may help to design synthetic networks with controlled dynamics.

\section{Deterministic representation of the stochastic dynamics}\label{sec:Langevindet}

As the gene network evolves under Eq. \eqref{eq:Langevin}, the probability distribution of gene expression states transforms according to its associated FP equation $\partial_tp(q;t)=-\nabla_q\cdot J$, with $J$ given by \cite{villarreal_general_2012}
\begin{equation}
J_i=\{\Theta[w_i(q)]-\alpha_iq_i\}p(q;t)-\frac{1}{2}Q\frac{\partial p}{\partial q_i},
\end{equation}
where $Q$ denotes the noise intensity: $\langle\xi_i(t)\xi_j(t')\rangle=:Q\delta_{ij}\delta(t-t').$ The analytical solution of this FP equation is \cite{villarreal_general_2012}
\begin{equation}\label{eq:FPsol}
    p(q,t)=\prod_{i=1}^N\frac{1}{\sqrt{a_i(t)\pi}}\exp\left\{-\sum_{i=1}^N\frac{(q_i-\mu_i(t))^2}{a_i(t)}\right\},
\end{equation}
a multivariate Gaussian where the time dependence arises only through the parameters $\mu_i$ and $a_i$. In the adiabatic approximation\cite{villarreal_general_2012}, they relax to steady-state values along the curve given by
\begin{subequations}\label{eq:mua}
\begin{align}
\mu_i(t) &= \mu_i(0)e^{-\alpha_i t} + q_i^*\left(1 - e^{-\alpha_i t}\right),\label{eq:mu} \\
a_i(t) &= \frac{Q}{\alpha_i}\left(1 - e^{-2\alpha_i t}\right) + a_i(0)e^{-2\alpha_i t}, \label{eq:a}
\end{align}
\end{subequations}
where $q_i^* = \Theta[w_i(\langle q\rangle )]/\alpha_i$ and $a_i^*:=Q/\alpha_i$ characterize the steady state. We assume the dynamics governed by Eq. \eqref{eq:Langevin} occur near a basin of attraction within an \textit{epigenetic landscape}, a conceptual model that visualizes development processes as a ball rolling stochastically through a terrain of peaks and valleys toward attractors at basin minima \cite{waddington2014strategy}. Specifically, we consider either a landscape with a single attraction basin at $q^*$, or dynamics localized near one basin in a multi-basin landscape.

The solution of the FP equation translates the stochastic dynamics of the network into deterministic trajectories in a finite-dimensional space spanned by $(\mu_i,a_i)$ (see Fig. \ref{fig:stoc-to-det}). The corresponding flow is obtained by differentiating Eqs. \eqref{eq:mua} (see the Supplementary Material for explicit calculations):
\begin{subequations}\label{eq:dyn}
\begin{align}
\dot{\mu}_i &= -\alpha_i\left(\mu_i - q_i^*\right),\label{eq:dynm} \\
\dot{a}_i &= -2\alpha_i\left(a_i - \frac{Q}{\alpha_i}\right). \label{eq:dyna}
\end{align}
\end{subequations}

\begin{figure}
 \includegraphics[width=\textwidth]{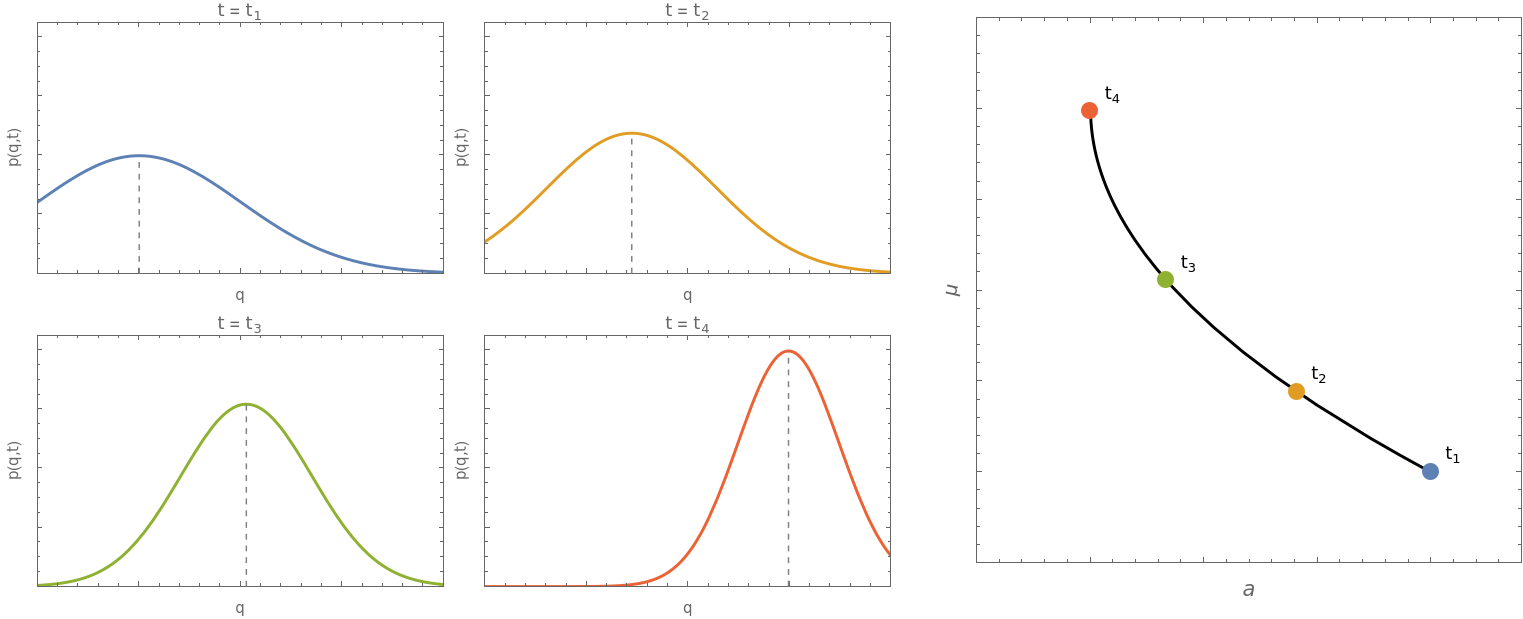}
 \caption{This illustrates how we represent the FP dynamics of the network with a deterministic one. The left panel depicts the evolution of $p(q;t)$ in time for one node of the network. Each one of these distributions corresponds to a point in parameter space. Thus, the solution to the corresponding Fokker-Planck equation can be regarded as a curve on this finite-dimensional space.}\label{fig:stoc-to-det}
\end{figure}

As the network converges toward a steady state, its current state grows increasingly ``similar'' to this final attractor in terms of their information content. This ``similarity'' is quantified by the \textit{Kullback-Leibler divergence}. Divergences are asymmetric ``distance-like'' measures between distributions. Unlike proper distances, they do not obey the triangle inequality, but they are non-negative and vanish only when the two distributions coincide.
In particular, the KL divergence is defined as
\begin{equation}\label{eq:KL}
D_\text{KL}(P_1(q)||P_2(q)) := \int_\mathcal{Q} P_1(q) \log\left(\frac{P_1(q)}{P_2(q)}\right) \mathrm{d}q,
\end{equation}
where integration runs over all possible network states $q\in\mathcal{Q}$, which we assume continuous.

Since the network obtains information from the steady state as it evolves in time, it is natural to wonder about dynamics that maximize the rate at which this information is gained.
Geometrically, this amounts to asking when Eq. \eqref{eq:dyn} acts as the gradient flow of the KL divergence.

To consider the gradient descent of the KL divergence, we need to pin down what we exactly mean by \textit{gradient}, for which two common notions exist. First, we have the so-called \textit{vanilla gradient} (vector of partial derivatives), which depends on how we parameterize the distributions. This is problematic because if we reparameterize, the new steepest descent direction might not coincide with that given by the natural parameters $(\mu_i,a_i)$. The second notion, the \textit{natural gradient}, avoids this pitfall,
since it is coordinate-independent.
Although its definition requires a metric tensor,
this is not a difficulty in this case,
since the KL divergence provides us with one \cite{amari_information_2016}.

To show this, we observe that the KL divergence induces a divergence on the space spanned by the parameters $\eta$ of \textit{any} family of distributions:
\begin{equation*}
D(\eta_1||\eta_2):=D_\text{KL}(p(q;\eta_1)||p(q;\eta_2)).
\end{equation*}
 This is well defined if we assume that the identification $\eta\mapsto p(q;\eta)$ is one-to-one.
 Furthermore, as we show in the SM,
 for nearby parameters,
\begin{equation}
 D(\eta||\eta+\delta\eta)\approx -\frac{1}{2}E_\eta\left[\frac{\partial^2\ell_q}{\partial\eta^i\partial\eta^j}\right]\delta\eta^i\delta\eta^j,
\end{equation}
with $\ell_q(\eta):=\log p(q;\eta)$ denoting the log-likelihood and 
$E_\eta[\cdot]$ meaning the expectation with respect to $p(q;\eta)$. Above, and in what follows,
we use Einstein's summation convention over repeated indices.
We show that, in particular,
for the family of distributions defined by Eq. \eqref{eq:FPsol}, where $\eta=(\mu_i,a_i)$ (see SM),
 \begin{equation*}
 \begin{split}
E_\eta\left[\frac{\partial^2\ell_q}{\partial\mu_i\partial\mu_j}(q;\eta)\right]& =-\frac{2}{a_i}\delta_{ij},\\
E_\eta\left[\frac{\partial^2\ell_q}{\partial\mu_i\partial a_j}(q;\eta)\right]& = 0,\\
E_\eta\left[\frac{\partial^2\ell_q}{\partial a_i\partial a_j}(q;\eta)\right]& = -\frac{\delta_{ij}}{2 a_ i^2}.
\end{split}
 \end{equation*}
 Considering that the $a_i>0$,
 the matrix with entries $g_{ij}:=-E_\eta[\partial_i\partial_j\ell_q]$ is positive definite,
and thus defines a metric tensor on the parameter space,
called \textit{Fisher metrics}.
We use it to compute the gradient flow of the KL divergence in the next section.

It is important to remark that the $g_{ij}$ above yield a positive definite matrix for \textit{any} parametric family of distributions \cite{amari_information_2016}.
In other words,
the parameter space of every such family is a Riemannian space with the Fisher metrics. This is one of the three ingredients of \textit{information geometry}.

\section{Dynamics of fastest divergence decrease}\label{sec:D1} 

For the Gaussian family in Eq. \eqref{eq:FPsol},
the KL divergence is (see SM)
\begin{equation}\label{eq:KL}
D((\mu_1,a_1)||(\mu_2,a_2)) = \sum_{i=1}^N \left\{ \frac{1}{2}\left[\frac{{a_i}_1}{{a_i}_2}-\log\left(\frac{{a_i}_1}{{a_i}_2}\right)-1\right] +\frac{({\mu_i}_1-{\mu_i}_2)^2}{{a_i}_2} \right\}.
\end{equation}

\begin{figure}
\centering
 \includegraphics[scale=0.5]{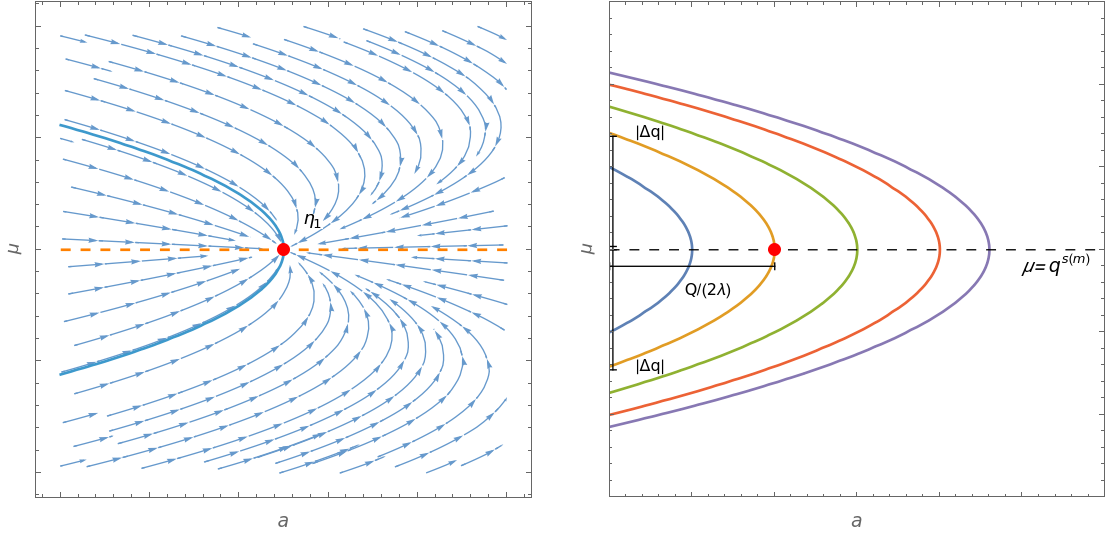}
 \caption{Left panel: The gradient flow of $D_1$ on a $a_i$-$\mu_i$ plane, for some fixed $i$. The solid curve and the dashed line are the only trajectories of this dynamics that also correspond to the FP dynamics of the node.
 Right panel: Greater noise allows for a greater differentiation between the initial and the final states of the network, if its evolution occurs along the gradient flow of $D_1$.}\label{fig:D1-nd}
\end{figure}

Fix $\eta_1:=\left(q_i^*,Q/\alpha_i\right)$. To compute the gradient of $D(\eta_1||\cdot)=D_1$, we apply the inverse Fisher metrics $g^{ij}$ to $\partial_jD_1$, obtaining the following \textit{Riemannian} gradient flow converging to $\eta_1$ \cite{lee_riemannian_1997}:
\begin{subequations}\label{eq:desgradD1}
\begin{align}
\dot{\mu}_i &= -\lambda \left(\mu_i -q_i^*\right), \label{eq:desgradD1mu} \\
\dot{a}_i &= -\lambda\left[a_i - \frac{Q}{\alpha_i} - 2\left(\mu_i -q_i^*\right)^2\right],\label{eq:desgradD1a}
\end{align}
\end{subequations}
where $\lambda>0$. This flow is represented in the left panel of Fig. \ref{fig:D1-nd} for one node of the network.

Comparing this with Eq. \eqref{eq:dyn}, the network evolves along the gradient flow of $D_1$ if
\begin{subequations}
\begin{align}
-\alpha_i(\mu_i -q_i^*) &= -\lambda (\mu_i -q_i^*),\\
-2\alpha_i\left(a_i - \frac{Q}{\alpha_i}\right) &= -\lambda\left[a_i - \frac{Q}{\alpha_i} - 2\left(\mu_i -q_i^*\right)^2\right].
\end{align}
\end{subequations}
These equations constrain the decay rates $\alpha_i$ and steady-state values $\left(q_i^*,Q/\alpha_i\right)$. To see how, we consider three separate cases. \textit{i)} First, observe that fixed variances ($a_i(t)\equiv Q/\alpha_i$) imply constant means. So, maximizing the rate at which $D_1$ decreases requires non-constant variances. \textit{ii)} Second, if the means are fixed ($\mu_i(t)\equiv q_i^* $), all $a_i$ must decay uniformly at a rate $\alpha_i\equiv\lambda/2$, as follows from Eq. \eqref{eq:desgradD1a}. \textit{iii)} Third, when the $\mu_i$ are non-constant, they also decay uniformly, but at a rate $\alpha_i\equiv\lambda$ along the curve
    \begin{equation}\label{eq:relaxcurve}
    a_i - \frac{Q}{\alpha_i} = -2\left(\mu_i -q_i^*\right)^2.
    \end{equation}

For one node in the network, this curve corresponds to the solid parabola in the left panel of Fig. \ref{fig:D1-nd}. In this case, the variances also decay uniformly. Furthermore, if the initial $a_i(0)$ is close to zero, the $a_i(t)$ correspond to the mean square fluctuations of $q_i$, \textit{i. e.}, $a_i=\langle q_i^2(t)\rangle-\langle q_i(t)^2\rangle$ \cite{villarreal_general_2012}. This leads to a relationship between the noise intensity and the parameters of the dynamics that resembles a fluctuation dissipation theorem, where the noise intensity $Q$ is approximately proportional to the product of the decay rate $\lambda$ (analogous to a damping constant) and the square of the transition distance:
\begin{equation}
Q \approx 2\lambda \left(\mu_i(0) -q_i^*\right)^2,
\end{equation}
as follows after substituting Eq. \eqref{eq:mua} into the relaxation curve \eqref{eq:relaxcurve}. This result suggests that in the regime of small uncertainties, a larger differentiation is inherently tied to higher levels of stochastic fluctuations in the dynamics. Note that this result is observed in various genetic processes such as transcription, coordinated gene expression, signaling pathways, and cell fate \cite{eldar_functional_2010,gomez2020multidimensional,azpeitia2020signaling,pal2024living,tkacik_information_2025}. In the right panel of Fig. \ref{fig:D1-nd}, we illustrate how noise allows for a greater difference between the initial and final expression levels of one node.

To close this section, we recall that the function $D_1$ measures how inefficient it is, in terms of information, to code samples of $p(q;\eta_1)$ using a different distribution \cite{cover_elements_2006}. In other words, it tells us how much extra information the network currently has, when compared to its steady state. Our first finding is that, to lose this information excess the fastest, all variances must decay uniformly.

\section{Dynamics of fastest information gain.}\label{sec:D1*} 

\begin{figure}
\centering
 \includegraphics[scale=0.5]{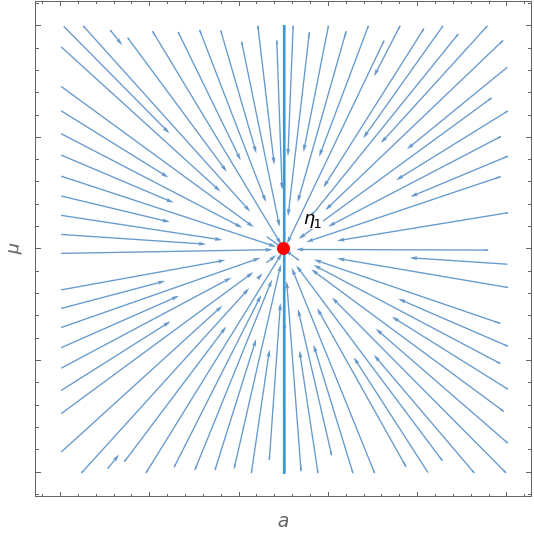}
 \caption{Gradient flow of $D_1^*$. The solid vertical line represents the trajectory of the only FP dynamics occuring along this flow.}\label{fig:reversedKL}
\end{figure}

Recall that divergences are nonsymmetric. Therefore, dynamics that maximize the rate of information efficiency will not be the same as those that maximize the rate of information gain, which correspond to the gradient flow of the dual divergence $D(\cdot||\eta_1)=:D_1^*$, given by
\begin{subequations}\label{eq:desgrad2}
    \begin{align}
        \dot\mu&=-\lambda\frac{a_i\alpha_i}{Q}\left(\mu_i-q_i^*\right)\label{eq:desgrad2mu},\\
    \dot a_i&=-\lambda\frac{a_i\alpha_i}{Q}\left(a_i-\frac{Q}{\alpha_i}\right)\label{eq:desgrad2a}.
    \end{align}
\end{subequations}
See SM for details. In Fig. \ref{fig:reversedKL} we show the gradient flow of the dual divergence for one node. The solid line represents the only solution to the FP equation along this flow, which we obtain by matching Eqs. \eqref{eq:desgrad2} with Eqs. \eqref{eq:dyn}:
\begin{subequations}
    \begin{align}
        -\alpha_i\left(\mu_i-q_i^*\right)    &=-\lambda\frac{a_i\alpha_i}{Q}\left(\mu_i-q_i^*\right),\\
        -2\alpha_i\left(a_i-\frac{Q}{\alpha_i}\right)&=-\lambda\frac{a_i\alpha_i}{Q}\left(a_i-\frac{Q}{\alpha_i}\right).
    \end{align}
\end{subequations}
To satisfy these equations,
the variances $a_i$ must remain fixed at $Q/\alpha_i$.
This forces all means to decay uniformly at a rate $\alpha_i\equiv\lambda$. 

What does the above tell us about the evolution of the network? Maximizing  the information gain requires that ``uncertainty'' in the expression of each gene is kept constant as the system reaches equilibrium. At the same time, all genes in the network must adjust their expression levels in a coordinated fashion, \textit{i. e.,} no gene may dominate over the others, as all the decay rates are the same. So, the system acquires the most information from the steady states, not from decreasing its uncertainty, but from suppressing all prominent gene expressions. This result is consistent with the optimization of self-organized biological systems, the population-level responses due to a single cell, and the transcriptional regulators in ligand-receptor circuits\cite{eldar_functional_2010,azpeitia2020signaling,tkacik_information_2025,pal2024living}.

\section{Dynamics along geodesics}\label{sec:geo} 

Viewing the dynamics of a GRN through the lens of information theory reveals a key property about the paths that optimize information: they may be described as geodesic curves. To see this, let $\eta_0$ and $\eta_1$ denote the initial and final states of the gene network, and consider the interpolation between them via an exponential mixture:
\begin{equation}
p(q;t)=\frac{p(q;\eta_1)^tp(q;\eta_0)^{1-t}}{C(t)},
\end{equation}
 where $C(t)$ is a normalization constant and $0\leq t\leq1$.
If this mixture lies within our parametric family (Eq. \eqref{eq:FPsol}),
it traces a path in parameter space. To find this curve,
denoted by $\eta(t)$,
take the logarithm and then differentiate with respect to $t$:
\begin{equation}
    \frac{\partial\ell_q}{\partial \eta^i}\dot\eta^i=\ell_q(\eta_1)-\ell_q(\eta_0)-   \frac{\mathrm{d}}{\mathrm{d}t}\log C(t).
\end{equation}
Notice that we used the Chain Rule on $\dot\ell_q$. We differentiate again and take expectations. This gives
\begin{equation}
g_{ij}\dot\eta^i\dot\eta^j=\frac{\mathrm{d}^2}{\mathrm{d}t^2}\log C(t),
\end{equation}
using that $E_\eta[\partial_i\ell_q]=0$.
This result relates the (Riemannian) speed of the curve to the normalization constant, telling us that $\log C$ is a convex function. So, regardless of whether
\begin{equation}
 C(t)=\int_\mathcal{Q}p(q;\eta_1)^tp(q;\eta_0)^{1-t}\mathrm{d}q,
\end{equation}
increases or decreases, it will attain a minimum, whose negative is known as the \textit{Chernoff distance} between the initial and final distribution, used in hypothesis testing \cite{cover_elements_2006}. If in the previous step we multiply by $\partial_k\ell_q$ before taking expectations,
what we obtain is
\begin{equation}\label{eq:geoexp}
 g_{ki}\ddot\eta^i+E_\eta\left[\frac{\partial\ell_q}{\partial\eta^k}\frac{\partial^2\ell_q}{\partial\eta^i\partial\eta^j}\right]\dot\eta^i\dot\eta^j=0,
\end{equation}
resorting to an alternative expression for the Fisher metrics: $g_{ij}=E_\eta[\partial_i\ell_q\partial_j\ell_q]$ \cite{amari_information_2016}.

The equation above represents a \textit{geodesic},
a curve of constant velocity as measured by a \textit{connection}. This
geometric object embodies a parameter-independent notion of ``straightness,'' and is locally determined by a set of $(2N)^3$ functions known as \textit{Christoffel symbols}. For more on connections, see SM.

Eq. \eqref{eq:geoexp} means that, when the exponential interpolation of two distributions of the same family traces a curve on the parameter space, this curve is \textit{straight} with respect to the so-called \textit{exponential connection}, whose Christoffel symbols are
\begin{equation}
\Gamma^*_{ij,k}:=E_\eta\left[\frac{\partial\ell_q}{\partial\eta^k}\frac{\partial^2\ell_q}{\partial\eta^i\partial\eta^j}\right].
\end{equation}
Like the Fisher metrics, such a connection can be defined on the parameter space of any family of distributions. This is the second central object in information geometry.

In contrast to geodesics, relaxations to a steady state do not occur at constant speed, since equilibrium is reached asymptotically. If we allow for the speed of the geodesics above to decrease by changing the right-hand side to $-\lambda g_{ki}\dot\eta^i$, which amounts to a reparameterization of the curve,
Eqs. \eqref{eq:geoexp} match Eqs. \eqref{eq:desgrad2} (see SM).
This tells us that the gradient descent of $D_1^*$ occurs along a \textit{pregeodesic}  of the exponential connection converging to $\eta_1$.
In other words, relaxations maximizing the rate at which the system gains information from the steady state are exponential combinations of the initial and final states of the network. This is a key result of our work and is consistent with the results reported by some authors\cite{tkacik_information_2025,momin_unlocking_2025}, in which stochastic fluctuations determine the fate of cells.

For the divergence $D_1$, the analogous interpolation is linear: $ p(q;t)=p(q;\eta_0)(1-t)+p(q;\eta_1)(t)$, with $0\leq t\leq1$. Repeating the differentiation process yields
 \begin{equation}
 g_{ki}\ddot\eta^i+E_\eta\left[\dfrac{\partial\ell_q}{\partial\eta^k}\left(\dfrac{\partial\ell_q}{\partial\eta^i}\dfrac{\partial\ell_q}{\partial\eta^j}+\dfrac{\partial^2\ell_q}{\partial\eta^i\partial\eta^j}\right)\right]\dot\eta^i\dot\eta^j=0.
 \end{equation}

Like the case of exponential mixtures, the functions
 \begin{equation}
 \Gamma_{ij,k}:=E_\eta\left[\dfrac{\partial\ell_q}{\partial\eta^k}\left(\dfrac{\partial\ell_q}{\partial\eta^i}\dfrac{\partial\ell_q}{\partial\eta^j}+\dfrac{\partial^2\ell_q}{\partial\eta^i\partial\eta^j}\right)\right],
 \end{equation}
 define a connection, known as \textit{mixture connection}.
The gradient descent of $D_1$ follows its geodesics at nonconstant speed (again,
the steady state is reached asymptotically).
Thus,
relaxations that are linear interpolations of the initial and the final state of the network maximize the rate at which it dumps the information that exceeds the one provided by the steady state.

So far, we have not considered \textit{metric} geodesics.
Since we can also determine directions with Riemannian tensors,
there is a concept of ``straightness'' associated to them.
The connection capturing this is called \textit{Levi-Civita} or \textit{metric connection}.
Its Christoffel symbols are obtained from the metric \cite{lee_riemannian_1997}:
\begin{equation}
 \Gamma^g_{ij,k}=\frac{1}{2}\left(\frac{\partial g_{jk}}{\partial \eta^i}+\frac{\partial g_{ik}}{\partial \eta^j}-\frac{\partial g_{ij}}{\partial \eta^k}\right).
\end{equation}

\begin{figure}
\centering
 \includegraphics[scale=0.45]{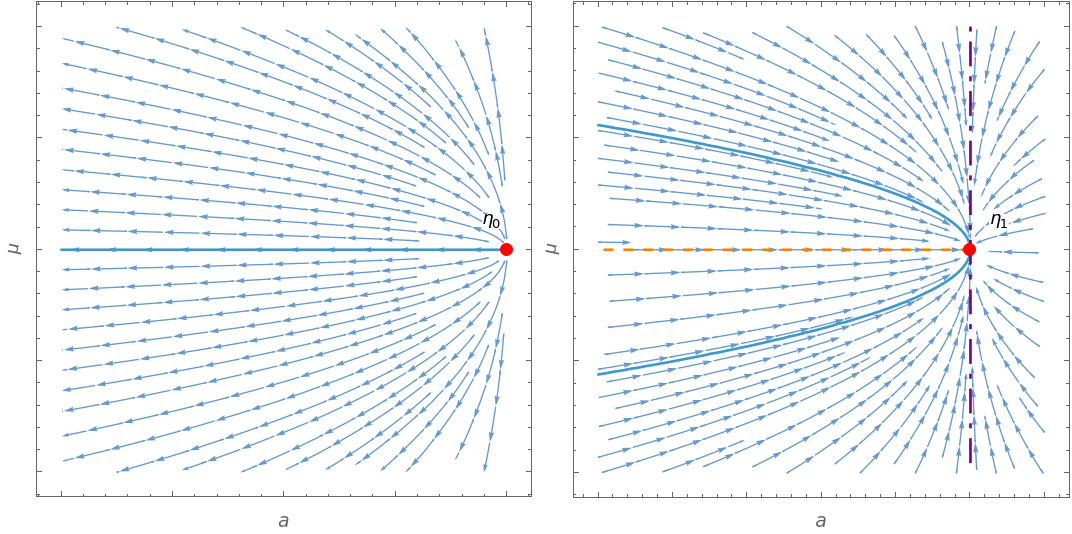}
 \caption{Left: Flow of Fisher geodesics originating at $\eta_0$. Notice that the geodesics are ellipses. The solid, blue horizontal line is the only one corresponding to a trajectory of the FP dynamics. Right: As a graphical summary of our work, we show the FP dynamics converging to $\eta_1$ for one node. The solid curve is a mixture geodesic, the dot-dashed line is an exponential geodesic and the dashed line is a Fisher geodesic running in the opposite direction: it starts at $\eta_1$ and converges to $a=0$. The trajectories of these dynamics are parabolae: despite looking very similar to Fisher geodesics, they only coincide when they degenerate to a horizontal line. }\label{fig:Fisher-Langevin}
\end{figure}

Under certain conditions,
metric geodesics define shortest paths in Riemannian geometry.
So,
it makes sense to ask if the dynamics in Eqs. \eqref{eq:dyn} ever minimize the Fisher distance between the initial and the final state of the network.
To answer to this question,
we compare the geodesic equations for the Levi-Civita connection, which we compute in SM and illustrate in the left panel of Fig. \ref{fig:Fisher-Langevin},
 with the relaxation dynamics on the parameter space.
 We do this by differentiating Eqs. \eqref{eq:dyn} and plugging the result in the geodesic equations for the Levi-Civita connection.
 This yields
\begin{subequations}\label{eq:LCgeo}
    \begin{align}
        -\alpha_i\dot\mu_i-\frac{\dot\mu_i\dot a_i}{a_i}&=-\lambda\dot\mu_i\\
        -2\alpha_i\dot a_i+2\dot\mu_i^2-\frac{\dot a_i^2}{a_i}&=-\lambda\dot a_i.
    \end{align}
\end{subequations}
In this case,
the only relevant dynamics takes place at constant means. This is because constant variances along a Fisher geodesic imply constant means, and nonconstant $\mu_i$ and $a_i$ cannot occur along this flow since such dynamics require that $\dot\mu_i^2=-\alpha_ia_i(0)$,
which contradicts the fact that all the $a_i$ are positive.
The unique Fisher geodesic that is also a trajectory of Eq. \eqref{eq:Langevin} is represented by a solid line on the left panel of Fig. \ref{fig:Fisher-Langevin}. 

Furthermore, observe that if $\dot\mu_i=0$, then $\lambda-2\alpha_i-\dot a_i/a_i=0$. Solving for $a_i$ forces $\lambda$ and $Q$ to be both zero,
comparing to Eq. \eqref{eq:a}.
Thus,
only deterministic dynamics develop along Fisher geodesics.
These paths approach the boundary $a_i \to 0$ asymptotically at constant speed: as the network evolves, the uncertainty about the expression level of each gene vanishes. This dynamics lacks any biological significance, since noise, a fundamental ingredient of gene expression, is suppressed.

Kim \cite{kim_information_2021} identifies Fisher geodesics as those minimizing the cumulative biological cost of changing the state of a system. So, according to our restult, to minimize the biological cost of changing the state of a gene network, one must switch noise off, which results in undifferentiated states. In other words, some nonminimal biological cost is necessary for cell differentiation in the context of Eq. \eqref{eq:Langevin}. We present a graphical summary of this section in the right panel of Fig. \ref{fig:Fisher-Langevin}. This figure shows the flow of the FP dynamics for one node of the network. The three superimposed curves are trajectories of this flow and geodesics at the same time. The solid curve corresponds to a mixture geodesic, the dot-dashed vertical line is an exponential geodesic, and the dashed horizontal line is a metric geodesic.

\section{Concluding remarks and perspectives}

Our informational study of the dynamics of GRNs provides interesting insights about these systems. When their evolution occurs along a geodesic, we can interpret the decay rates $\alpha_i$ as damping constants. Following this analogy, the most efficient dynamics, in terms of information, take place when this damping force is isotropic, \textit{i. e.}, all the $\alpha_i$ are the same. In this situation, noise and uncertainty in gene expression levels become the central characters.

We showed that the system optimizes informational efficiency as it evolves towards equilibrium through the evolution of its uncertainties. When uncertainties are small, noise levels influence the potential differentiation between initial and final states.

On the other hand, when the uncertainties in gene expression are kept constant, the dynamics maximize the rate at which the system gains information from the steady state. Here, the noise does not drive differentiation, but balances the decay rate to maintain these uncertainties at a fixed value.

We also showed that, in the regime where Eq. \eqref{eq:Langevin} holds, evolution with the minimum biological cost requires the suppression of noise. Here, the system evolves to a concentrated state with no uncertainty about the level of expression of its genes. This may be interpreted as noise driving cell fate determination, coinciding with similar existing interpretations \cite{hledik_accumulation_2022,pal2024living,momin_unlocking_2025} with experimental support \cite{eldar_functional_2010,azpeitia2020signaling}.

The approach we adopted here also allows us to portray noise and decay rates as controls  that can be tuned to drive the evolution of a network along a desired trajectory, which could be optimal in terms of information \cite{kim_geometric_2016} or helpful in drug design \cite{karolak_concepts_2021, momin_unlocking_2025}.

The central role of isotropic noise in enabling efficient transitions invites an intriguing extension to population-level models. Our geometric framework could be used to explore whether a principle analogous to Fisher's fundamental theorem \cite{baez_fundamental_2021}, where genetic diversity governs the rate of adaptive evolution, manifests in the context of cell populations navigating an epigenetic landscape. Specifically, one could investigate if the variance in gene expression states across a population predicts the efficiency of collective adaptation to new environmental attractors.

Another interesting direction for future work would be to model asymmetric relaxations in networks, where KL-equidistant initial states relax to equilibrium at different rates. This phenomenon is well documented in non-equilibrium systems \cite{ibanez_heating_2024}, and has recently been addressed using the same information-geometric framework employed here \cite{bravetti_asymmetric_2025}. Specifically, the role of the Amari-Chentsov tensor in predicting relaxation asymmetries could be explored within our model of GRN dynamics. This might reveal new principles underlying the fundamental differences between differentiation and reprogramming pathways.

The Fisher information $I(t):=g_{ij}\dot\eta^i\dot\eta^j$ which quantifies the speed of statistical evolution, decreases monotonically as a network relaxes towards equilibrium. This decay implies  a reduction of the entropy production rate $\sigma^{\text{tot}}(t)$, which itself vanishes as the system reaches equilibrium \cite{ito_stochastic_2020}. Future work could explore non-equilibrium steady states where entropy production remains finite despite vanishing Fisher information, potentially revealing deeper connections between information and dissipation in the context of GRNs.

Finally, we remark that our work considers only transitions to a fixed attraction basin, as a first approach to address the informational evolution of a network across the full epigenetic landscape from a geometric point of view in the future.

\section*{Acknowledgements}
The work of MAGA was funded by SECIHTI, programa Estancias
Posdoctorales por M\'exico. JRRA acknowledges the support of DGAPA-UNAM under the program PAPIIT, Grant No. IA-109525.

\appendix

\section{Deterministic representation of stochastic dynamics and derivation of the Fisher metrics}\label{supp:fisher} 

This section provides detailed derivations and explicit calculations supporting the results in the main article. Eqs.\eqref{eq:dyn} represent the time evolution of the transition probability of the states of the gene network, and are obtained from Eqs. \eqref{eq:mua} by differentiating:
\begin{equation*} 
  \dot\mu_i =-\alpha_i(\mu_i(0)-q_i^*)e^{-\alpha_i t};~~~
  \dot a_i=-2\alpha_i\left(a_i(0)-\frac{Q}{\alpha_i}\right)e^{2\alpha_it},
 \end{equation*}
and then eliminating the parameter $t$. We do this solving for $e^{\alpha_i t}$ and $e^{-2\alpha_it}$ in Eqs. \eqref{eq:mua} and substituting in the eqs. above
\begin{align*}
        \dot\mu_i&=-\alpha_i(\mu_i(0)-q_i^*)\frac{\mu_i-q_i^*}{\mu_i(0)-q_i^*}=-\alpha_i\left(\mu_i-q_i^*\right);\\     
        \dot a_i&=-2\alpha_i\left(a_i(0)-\frac{Q}{\alpha_i}\right)\frac{a_i-\frac{Q}{\alpha_i}}{a_i(0)-\frac{Q}{\alpha_i}}=-2\alpha_i\left(a_i-\frac{Q}{\alpha_i}\right).
\end{align*}

We assert that the parameter space spanned by $(\mu_i,a_i)$ is a Riemannian manifold. Actually, the result is more general: the parameter space of \textit{any} family of distributions has a metric tensor, intuitively, the ``infinitesimal version'' of the KL divergence. Indeed, consider an expansion of $D_1$ to second order for $\eta=\eta_1+\delta\eta$.
The first-order coefficient is
\begin{align*}
 \left.\frac{\partial D_1}{\partial\eta^j}\right\vert_{\eta=\eta_1} & = \left.\frac{\partial}{\partial\eta^j}\right\vert_{\eta=\eta_1}\int_\mathcal{Q}p_1\log\frac{p_1}{p_\eta}\mathrm{d}q= -\int_\mathcal{Q}\left.\frac{\partial p_\eta}{\partial\eta^j}\right\vert_{\eta=\eta_1}\mathrm{d}q=0.
\end{align*}
This means that changes in the parameter do not affect the KL divergence to first order.
As usual, we are assuming that we can differentiate over the integral,
and we are denoting $p(\cdot;\eta)$ by $p_\eta$. If we differentiate again,
\begin{align*}
\frac{\partial^2D_1}{\partial\eta^i\partial\eta^j}&=-\frac{\partial}{\partial\eta^i}\int_\mathcal{Q}\frac{p_1}{p_\eta}\frac{\partial p_\eta}{\partial\eta^j}\mathrm{d}q=-\frac{\partial}{\partial\eta^i}\int_\mathcal{Q}p_1\frac{\partial\ell_q}{\partial\eta^j}\mathrm{d}q=-E_1\left[\frac{\partial^2\ell_q}{\partial\eta^i\partial\eta^j}\right].
\end{align*}

The coefficients $g_{ij}(\eta_1):=\left.\partial_i\partial_j D_1\right\vert_{\eta=\eta_1}$ define a positive definite matrix, regardless of the values of the $\eta_1$ and of the family of distributions \cite{amari_information_2016}. They are the components of the \textit{Fisher metrics} defined on the parameter space. 

We can also express the $g_{ij}$ in terms of the covariance matrix of the $\partial_j\ell_q$ by differentiating the product $1/p_\eta\partial p_\eta/\partial\eta^j$ instead of grouping it as $\partial\ell_q/\partial\eta_j$:
\begin{align*}
\left.\frac{\partial^2D_1}{\partial\eta^i\partial\eta^j}\right\vert_{\eta=\eta_1} &=-\int_\mathcal{Q}p_1\left.\left(-\frac{1}{p_\eta^2}\frac{\partial p_\eta}{\partial\eta^i}\frac{\partial p_\eta}{\partial\eta^j}+\frac{1}{p_\eta}\frac{\partial^2p_\eta}{\partial\eta^i\partial\eta^j}\right)\right\vert_{\eta=\eta_1}\mathrm{d}q=E_1\left[\left.\left(\frac{\partial\ell_q}{\partial\eta^i}\frac{\partial\ell_q}{\partial\eta^j}\right)\right\vert_{\eta=\eta_1}\right].
\end{align*}
The second term inside the integral is zero because it is just the second derivative of $\int_\mathcal{Q}p_\eta\mathrm{d}q=1$ at $\eta=\eta_1$. 

So, $D_1(\eta_1+\delta\eta)\approx\frac{1}{2}g_{ij}(\eta_1)\delta\eta^i\delta\eta^j$, which shows that the Fisher metrics is the ``infinitesimal version'' of the KL divergence. 

In the particular case of the family defined by Eq. \eqref{eq:FPsol}, we can write the Fisher metrics as the sum of the metric of each node of the network, provided that $\ell_q$ is a sum of logarithms of Gaussian distributions. In other words, we may write the Fisher metrics as a diagonal block matrix, each of which is the Fisher metrics of a normal family \cite{amari_information_2016}:
\begin{equation*}
 (g_{ij})=\left(
            \begin{array}{cc}
              2/a & 0\\
              0 & 1/(2a^2)
            \end{array}
            \right).
\end{equation*}

\section{Explicit calculations of gradient flows} 

Likewise, to obtain the KL divergence given by Eq. \eqref{eq:KL} we only need the KL divergence between two normal distributions \cite{calin_geometric_2014}:
\begin{equation*}
 D_\text{KL}(p(q;\mu_1,a_1)||p(q;\mu_2,a_2))=\frac{1}{2}\left(\frac{a_1}{a_2}-\log\frac{a_1}{a_2}-1\right)+\frac{(\mu_1-\mu_2)^2}{a_2}.
\end{equation*}
In what follows, we only do the explicit computations for one node.

We start with the gradient of the KL divergence. Recall that the gradient of a function $f$ is the vector field associated to the differential of the function via the metric tensor: $g(\operatorname{grad}f,\cdot )=\mathrm{d}f$, or in matrix notation, $(\operatorname{grad}f^i)^T(g_{ij})=(\partial_jf)$. The components of the gradient are thus obtained by multiplying $(\partial_jf)$ with the inverse of $(g_{ij})$, this is, $(\operatorname{grad}f^i)=[(\partial_jf)(g^{ij})]^T$. The gradient of
\begin{equation*}
D_1(\eta):=D(\eta_1||\eta)= \frac{1}{2}\left(\frac{a_1}{a}-\log\frac{a_1}{a}-1\right)+\frac{(\mu_1-\mu)^2}{a}
\end{equation*}
is then
\begin{equation*}
 (\operatorname{grad}D_1^i)=\left(\begin{array}{cc}
                                   a/2 & 0\\
                                   0 & 2a^2
                                  \end{array}
                                  \right) \left(\begin{array}{c}
                                           -2(\mu_1-\mu)/a\\
                                          (a-a_1-2(\mu_1-\mu)^2)/(2a^2)\\
                                          \end{array}
                                    \right)=\left(\begin{array}{c}
                                                   \mu-\mu_1\\
                                                   a-a_1-2(\mu-\mu_1)^2
                                                  \end{array}
                                            \right).
\end{equation*}
The components of the gradient flow $\dot\eta=-\lambda\operatorname{grad}{D_1}$ are $\dot\mu=-\lambda(\mu-\mu_1)$ and $\dot a=-\lambda[a-a_1-2(\mu-\mu_1)^2]$. 

We follow a similar procedure for
\begin{equation*}
 D_1^*(\eta):=D(\eta||\eta_1)=\frac{1}{2}\left(\frac{a}{a_1}-\log\frac{a}{a_1}-1\right)+\frac{(\mu-\mu_1)^2}{a_1},
\end{equation*}
whose gradient is
\begin{equation*}
 (\operatorname{grad}{D_1^*}^i)=\left(\begin{array}{cc}
                                   a/2 & 0\\
                                   0 & 2a^2
                                  \end{array}
                                  \right) \left(\begin{array}{c}
                                           2(\mu-\mu_1)/a_1\\
                                          1/2(1/a_1-1/a)\\
                                          \end{array}
                                    \right)=a/a_1\left(\begin{array}{c}
                                                   \mu-\mu_1\\
                                                   a-a_1
                                                  \end{array}
                                            \right).
\end{equation*}

\section{Affine connections and their geodesics.}

There is a parameter-independent notion of straight lines, called \textit{geodesics}. They are roughly defined as ``straight'' trajectories, \textit{i. e.},
curves whose tangent vector is always \textit{parallel} to the curve.
So,
defining geodesics relies on a definition of parallelism,
realized by what is known as an \textit{affine connection}.

Let us begin by explaining what affine connections are.
Suppose that the vectors $\{e_i\}_{i=1}^{M}$ span all possible independent directions in which we can move starting from a point $\eta$ in parameter space.
These $M$ vectors change from point to point, meaning that each $e_i$ is a vector field and at each point $\eta$. The $e_i$ are always linearly independent, whence we call $\{e_i\}$ a \textit{frame}. 

Taking this into account,
pick a vector $e_i$ and draw a curve that starts at $\eta$ and is always tangent to $e_i$ as we move from point to point.
After an infinitesimal displacement from $\eta$ to $\eta+\delta\eta$, a vector $e_j$ might not have the same direction (relative to the curve) as it did at the beginning.
``To first order'', intuitively speaking, the change in this vector will be $ e'_j=e_j+\Gamma_{ij}^ke_k$. The $\Gamma_{ij}^k$ are coeffients that measure how much we must correct on the $k$-th direction to recover the original direction of $e_j$, as measured with respect to $e_i$.
Clearly,
when the $\Gamma_{ij}^k$ are all zero,
the vector $e_j$ does not change along this curve,
and we can say that $e_j$ is \textit{parallel} to \textit{itself} along the direction of $e_i$.
Of course,
the correction coefficients $\Gamma_{ij}^k$ might change from point to point, so they are functions of $\eta$.

The conclusion of this is that we may define a notion of parallelism by means of $M^3$ real functions of the parameters.
This is called \textit{affine connection}.
The coefficients are the \textit{Ricci rotation coefficients} of the connection,
which are also called \textit{Christoffel symbols} when each vector $e_i$ is tangent to the curve $\eta^i=\text{const.}$ 

What if we wish to determine when a vector $X=X^je_j$ is parallel to itself along a curve tangent to $e_i$?
In this case,
we must also take into account the change in the components of $X$ along the curve, $\mathrm{d}X^j(e_i)$.
So, the total change of the vector as we move from $\eta$ to $\eta+\delta\eta$ is $\mathrm{d}X^j(e_i)e_j+X^k\Gamma_{ik}^je_j$.
This is what we call the \textit{covariant derivative} of $X$ with respect to the vector $e_i$ at $\eta$, which we denote by $(\nabla_iX)_\eta$. 
Usually, we refer to $\nabla$ as \textit{the connection}, and to the $\Gamma_{ij}^k$ as its Ricci coefficients (or Christoffel symbols) relative to the frame $\{e_i\}$ (or to the parameters $\eta$). 

We can extend covariant differentiation to arbitrary directions $Y$. We only need to add the contribution of each $\nabla_i X$ componentwise, \textit{i. e.}, $\nabla_YX=Y^j\nabla_jX$. 

It is important to point out that not any $M^3$ functions define a connection through its Christoffel symbols. For this, under a reparameterization $\eta\mapsto\theta$, the Christoffel symbols must change according to
\begin{equation*}
 \Gamma^k_{ij}\mapsto{\Gamma'}_{ij}^k=\frac{\partial\theta^k}{\partial\eta^l}\frac{\partial\eta^m}{\partial\theta^i}\frac{\partial\eta^n}{\partial\theta^j}\Gamma_{mn}^l+\frac{\partial\theta^k}{\partial\eta^l}\frac{\partial^2\eta^l}{\partial\theta^i\partial\theta^j}.
\end{equation*}
It is not hard to see that the Christoffel symbols of the exponential and mixture connections transform correspondingly, and therefore define connections on parameter space. 

Equipped with a connection, we can now pin down the concept of geodesic that we introduced heuristically before. We say that $\eta(t)$ is a geodesic if its tangent vector does not change along the curve. In components, this is $\ddot\eta^i+\Gamma_{jk}^i\dot\eta^j\dot\eta^k=0$. Multiplying both-hand sides by $g_{il}$ yields an equation in the form of Eq. \eqref{eq:geoexp}, defining $\Gamma_{ij,k}:=g_{kl}\Gamma_{ij}^l$.

\section{The gradient descents of the KL divergence are geodesics.}

Geodesics do not fully capture the idea of ``straight line.'' For instance, the curve $(x,y)=e^t(x_0,y_0)$ is still a straight line in the Euclidean plane, since its tangent vector changes magnitude, but remains parallel to itself along the curve. This more general notion is that of a \textit{pregeodesic}: a curve whose tangent vector does not change \textit{direction}. In this case, the covariant derivative of $\dot\eta$ along $\dot\eta$ is not zero, but parallel to $\dot\eta$ itself, this is, $\nabla_{\dot\eta}\dot\eta=-\lambda\dot\eta$.

In the particular case of normal distributions, the Christoffel symbols of the expontial connection are
\begin{equation*}
 (\Gamma^*_{ij,1})=\left(\begin{array}{cc}
                           0 & -2/a^2\\
                           -2/a^2 & 0
                          \end{array}
                      \right);~~~
  (\Gamma^*_{ij,2})=\left(\begin{array}{cc}
                            0 & 0\\
                            0 & -1/a^3
                          \end{array}
                    \right).
\end{equation*}
which, after some simplification, yield the pregeodesic equations
\begin{equation*}
 \ddot\mu-\frac{2}{a}\dot\mu\dot a=-\lambda\dot\mu;~~~\ddot a-\frac{2}{a}{\dot a}^2=-\lambda\dot a.
\end{equation*}

To compare these with the gradient descent
of $D_1$, differentiate both-hand sides of the latter with respect to $t$:
\begin{equation*}
    \ddot\mu = -\lambda\left[\frac{\alpha\dot a}{Q}\left(\mu-q^{s(m)}\right)+\frac{\alpha a}{Q}\dot\mu\right]=\frac{\dot\mu\dot a}{a}-\lambda\frac{\alpha a}{Q}\dot\mu,
\end{equation*}
where we substituted Eq. \eqref{eq:desgrad2mu} in the first term. From the same Eq. \eqref{eq:desgrad2mu}, we have that ${Q}/{\alpha}=-{\lambda a^2}/{\dot a-\lambda a}$. Plugging this into $\ddot\mu$ and rearranging terms yields the first pregeodesic equation. 

We follow an analogous procedure for the $a$-component of the gradient descent of $D_1^*$. Differentiate Eq. \eqref{eq:desgrad2a} with respect to $t$:
\begin{equation*}
 \ddot a=-\lambda\frac{\alpha\dot a}{Q}\left(a-\frac{Q}{\alpha}\right)-\lambda\frac{\alpha a}{Q}\dot a.
\end{equation*}
After substituting Eq. \eqref{eq:desgrad2a} in the first term and $Q/\alpha$ in the second one, we obtain the second pregeodesic equation. 

Let us turn our attention to the mixture connection. Its only nonzero Christoffel symbol is $ \Gamma_{11,2}={2}/{a^2}$. Then, the corresponding pregeodesic equations are 
\begin{equation*}
    \ddot\mu=-\lambda\dot\mu;~~~\ddot a+4{\dot\mu}^2=-\lambda\dot a.
\end{equation*}  
To compare these with the gradient descent of $D_1$,  differentiate Eq. \eqref{eq:desgradD1mu}, which straightforwardly yields the the first pregeodesic equation. Now, differentiating \eqref{eq:desgradD1a} with respect to $t$, we obtain $\ddot a=-\lambda\dot a+4\lambda\left(\mu-q^{s(m)}\right)\dot\mu$ which yields the second pregeodesic equation after substituting Eq. \eqref{eq:desgradD1mu} in the second term.

Finally, to obtain the metric pregeodesics, we compute the Christoffel symbols of the Levi-Civita connection:
\begin{equation*}
 (\Gamma^g_{ij,1})=\left(\begin{array}{cc}
                          0 & -1/a^2\\
                          -1/a^2 & 0
                         \end{array}
\right);~~~(\Gamma^g_{ij,2})=\left(\begin{array}{cc}
                                  1/a^2 & 0 \\
                                  0 & -1/(2a^3)
                                 \end{array}
\right).
\end{equation*}
Then, the metric pregeodesics are, after some simplification,
\begin{equation*}
 \ddot\mu - \frac{\dot\mu\dot a}{a} =-\lambda\dot\mu;~~~\ddot a+2{\dot\mu}^2-\frac{{\dot a}^2}{a}=-\lambda\dot a.
\end{equation*}

To find the values of the parameters in the dynamics that yield metric geodesics, we differentiate Eqs. \eqref{eq:dyn} and substitute above. This yields Eqs. \eqref{eq:LCgeo}, which we may rewrite as
\begin{equation*}
 \dot\mu\left(\lambda-\alpha-\frac{\dot a}{a}\right) = 0; ~~~ \dot a\left(\lambda-2\alpha-\frac{\dot a}{a}\right)  = -2{\dot\mu}^2.
\end{equation*}
We claim that nontrivial dynamics occur only when $\dot\mu=0$. If it were otherwise, we obtain from the first equation that $\dot a/a=\lambda-\alpha$. We may solve for $a$, obtaining $a=a_0e^{(\lambda-\alpha)t}.$ Plugging this into the second equation yields $\alpha(\lambda-\alpha)a=2{\dot\mu}^2$, 
implying that $(\lambda-\alpha)>0$. This would mean that $a$ diverges as $t\to\infty$, which contradicts the fact that the network relaxes to a steady state. When the mean is kept constant, a similar procedure gives $a=a_0e^{(\lambda-2\alpha)t}.$ We compare this to Eq. \eqref{eq:a}, obtaining that $\lambda=0$ and $Q=0$.


\end{document}